\documentclass[aps,floatfix,showpacs]{revtex4}
\usepackage{graphicx,amsfonts,amssymb,amsbsy}
\usepackage{pstricks}

\def\blue{\color{black}}

\usepackage{graphicx}
\usepackage{slashed}
\usepackage{bm}
\usepackage{amssymb}

\begin{document}

\title{\bf Deconfinement of neutron star matter within the Nambu-Jona-Lasinio model}
\author{G. Lugones$^1$, A.G. Grunfeld$^{2,3,4}$, N.N. Scoccola$^{2,3,5}$ and C. Villavicencio$^3$}

\affiliation{ $^1$ Universidade Federal do ABC, Centro de Ciencias
Naturais e Humanas, Rua Santa Ad\'elia, 166, 09210-170, Santo Andr\'e, Brazil\\
$^2$ CONICET, Rivadavia 1917, (1033) Buenos Aires, Argentina.\\
$^3$ Departmento de F\'\i sica, Comisi\'on Nacional de
Energ\'{\i}a At\'omica, (1429) Buenos Aires, Argentina.\\
$^4$ Department of Physics, Sultan Qaboos University, P.O.Box: 36
Al-Khode 123 Muscat, Sultanate of Oman
 \\
$^5$ Universidad Favaloro, Sol{\'\i}s 453, (1078) Buenos Aires,
Argentina.}

\vskip5mm
\begin{abstract}
We study the deconfinement transition of hadronic matter
into quark matter under neutron star conditions assuming color and
flavor conservation during the transition. We use a two-phase
description. For the hadronic phase we use different
parameterizations of a non-linear Walecka model which includes the
whole baryon octet. For the quark matter phase we use an $SU(3)_f$
Nambu-Jona-Lasinio effective model including color
superconductivity. Deconfinement is
considered to be a first order phase transition that conserves
color and flavor. It gives a short-lived transitory
colorless-quark-phase that is not in $\beta$-equilibrium, and
decays to a stable configuration in $\tau \sim \tau_{weak} \sim
10^{-8} s$. However, in spite of being very short lived, the transition to
this intermediate phase determines the onset of the transition inside neutron stars.
We find the transition free-energy density for temperatures typical of neutron
star interiors. We also find the critical mass above which compact stars
should contain a quark core and below which they are safe with respect
to a sudden transition to quark matter. Rather independently on the stiffness
of the hadronic equation of state (EOS) we find that the critical mass of
hadronic stars (without trapped neutrinos) is in the range of
$\sim$ 1.5 - 1.8  solar masses. This is in coincidence
with previous results obtained within the MIT Bag model.
\end{abstract}


\maketitle

\section{Introduction}

The core of compact stars can reach densities that are several
times larger than the saturation density of nuclear matter. In
such extreme conditions the baryons get so compressed that they
start to overlap and can produce a deconfined gas of quark matter.
An important characteristic of  the deconfinement transition in
neutron stars, is that just deconfined quark matter is
transitorily out of equilibrium with respect to weak interactions.
In fact, depending on the temperature, the transition should begin
with the quantum or thermal nucleation of a small quark-matter
drop near the center of the star. On the other hand, the flavor
composition of hadronic matter in  $\beta$-equilibrium is
different from that of $\beta$-stable quark-matter drop. Roughly speaking,
the direct formation of a $\beta$-stable quark-drop with $N$
quarks will need the almost simultaneous conversion of $\sim N/3$
{up and down} quarks into strange quarks, a process which is strongly
suppressed with respect to the formation of a non $\beta$-stable
drop by a factor $\sim G_{\mathrm{Fermi}}^{2N/ 3}$.  For typical
values of the critical-size $\beta$-stable drop ($N \sim 100-1000$
\cite{IidaSato1998}) the suppression factor is actually tiny.
Thus, quark flavor must be conserved during the deconfinement
transition
\cite{Madsen1994,IidaSato1998,Lugones1998,Lugones1999,Bombaci2004,Lugones2005,Bombaci2007}.
The main consequence of this condition is that the density of the
transition is higher than it would be if the direct formation of
$\beta$-stable quark-drops were possible. This is easy to
understand, since the Gibbs free energy per baryon of
$\beta$-equilibrated quark matter is always smaller of that of the
non-$\beta$-equilibrated state \cite{Lugones2005}.

Due to the uncertainties in
the knowledge of the state of matter at the densities of interest, studies
of the deconfinement transition are usually based on the extrapolation to
higher densities of an hadronic model valid around the nuclear saturation
density $\rho_{0}$, and the extrapolation to $\sim \rho_{0}$ of a quark model
that is expected to be valid only for high densities.
Within this kind of analysis the (in general) different functional
form of both EOSs, induces the phase transition to be first order.
Notice that from lattice QCD calculations there are indications
that the transition is actually first order in the high-density
and low-temperature regime, although this calculations involve
temperatures that are still larger than those in neutron stars,
and do not include the effect of color superconductivity
\cite{Fodor2004}.
In a previous work \cite{Lugones2005},  the deconfinement
transition has been analyzed within the frame of the MIT bag
model paying special attention to the role of color
superconductivity. In the present paper we shall analyze the
deconfinement transition employing the Nambu-Jona-Lasinio model in
the description of quark matter. For the hadronic phase we shall
use a model based on a relativistic Lagrangian of hadrons
interacting via the exchange of $\sigma$, $\rho$, and $\omega$
mesons \cite{GM1}. For simplicity, the analysis will be made in
bulk, i.e. without taking into account the energy cost due to
finite size effects in creating a drop of deconfined quark matter
in the hadronic environment.

This article is organized as follows. In Sec. \ref{hphase} we
briefly outline the non-linear Walecka model used to describe the
hadronic phase. In Sec. \ref{qphase} we provide some details of the
Nambu-Jona-Lasinio model we employ to describe the quark matter
phase paying special attention to the conditions of color
and flavor conservation. In Sec. \ref{disc} we present and discuss
our numerical results. Finally, in Sec. \ref{summary} a summary
with some conclusions is given.


\section{The hadronic phase}
\label{hphase}

For the hadronic phase we shall use a non-linear Walecka model
(NLWM) \cite{qhd,GM1,Menezes} which includes the whole baryon octet.
The Lagrangian of the model is given by
\begin{equation}
{\cal L}={\cal L}_{B}+{\cal L}_{M}+{\cal L}_{L},
\label{octetlag}
\end{equation}
where the indices $B$, $M$ and $L$ refer to baryons, mesons and
leptons respectively. For the baryons we have
\begin{eqnarray}
{\cal L}_B= \sum_B \bar \psi_B
\bigg[\gamma^\mu\left (i\partial_\mu -
g_{\omega B} \ \omega_\mu- g_{\rho B} \ \vec \tau \cdot \vec \rho_\mu \right)
-(m_B-g_{\sigma B} \ \sigma)\bigg]\psi_B,
\end{eqnarray}
with $B$ extending over the nucleons $N= n$, $p$ and the following
hyperons $H = \Lambda$, $\Sigma^{+}$, $\Sigma^{0}$, $\Sigma^{-}$,
$\Xi^{-}$, and $\Xi^{0}$. The contribution of the mesons $\sigma$,
$\omega$ and $\rho$ is given by
\begin{eqnarray}
{\cal L}_{M} &=& \frac{1}{2} (\partial_{\mu} \sigma \ \! \partial^{\mu}\sigma -m_\sigma^2 \ \! \sigma^2)
- \frac{b}{3} \ \! m_N\ \! (g_\sigma\sigma)^3 -\frac{c}{4} \ (g_\sigma \sigma)^4
\nonumber\\
& & -\frac{1}{4}\ \omega_{\mu\nu}\ \omega^{\mu\nu} +\frac{1}{2}\ m_\omega^2 \ \omega_{\mu}\ \omega^{\mu}
-\frac{1}{4}\ \vec \rho_{\mu\nu} \cdot \vec \rho\ \! ^{\mu\nu}+
\frac{1}{2}\ m_\rho^2\  \vec \rho_\mu \cdot \vec \rho\ \! ^\mu,
\end{eqnarray}
where the coupling constants are
\begin{eqnarray}
g_{\sigma B}=x_{\sigma B}~ g_\sigma,~~g_{\omega B}=x_{\omega B}~ g_{\omega},~~g_{\rho B}=x_{\rho B}~ g_{\rho}
\end{eqnarray}
where $x_{\sigma B}$, $x_{\omega B}$ and $x_{\rho B}$ are equal to $1$
for the nucleons and acquire different values for the other
baryons depending on the parametrization (see below).
The leptonic sector is included as a free Fermi gas which does not interact with the hadrons, i.e.

\begin{equation}
{\cal L}_{L}=\sum_l \bar \psi_l \left(i \rlap/\partial -
m_l\right)\psi_l.
\end{equation}


\begin{table*}[t!]
\centering
\begin{tabular}{l|c|c|c|c|c|c|c|c|cc}
\hline\hline
Label & composition & $x_{\sigma} = x_{\rho}$ & $x_{\omega}$ & $(g_{\sigma}/m_{\sigma})^2 $ & $(g_{\omega}/m_{\omega})^2$     &  $(g_{\rho}/m_{\rho})^2 $  &  $b$   &   $c$  &  $M_{max}$ \\
&  &  &  & $~\mathrm{[fm^2]}$ & $ ~\mathrm{[fm^2]}$     &  $ ~\mathrm{[fm^2]}$  &   &    &  \\
\hline
GM 1 & baryon octet + $e^-$  & 0.6 & 0.653    &  11.79  & 7.149 & 4.411   & 0.002947  & -0.001070 & 1.78 $M_{\odot}$ \\
GM 4 & baryon octet + $e^-$ & 0.9 & 0.9    &  11.79  & 7.149 & 4.411   & 0.002947  & -0.001070 & 2.2 $M_{\odot}$ \\
GM 5 & nucleons + $e^-$     & 0.6 & 0.653  & 11.79  & 7.149 & 4.411   & 0.002947  & -0.001070 & 2.35 $M_{\odot}$ \\
\hline\hline
\end{tabular}
\caption{Parameters of the hadronic equation of state. For each parametrization we give the maximum mass $M_{max}$
of a hadronic star.}
\label{setshadronicos}
\end{table*}


There are five constants in the model that are determined by the
properties of nuclear matter, three that determine the nucleon
couplings to the scalar, vector and vector-isovector mesons
$g_{\sigma}/m_{\sigma}$, $g_{\omega}/m_{\omega}$,
$g_{\rho}/m_{\rho}$, and two that determine the scalar self
interactions $b$ and $c$. It is assumed that all hyperons in the
octet have the same coupling than the $\Lambda$. These couplings
are expressed as a ratio to the nucleon couplings mentioned above,
that we thus simply denote $x_\sigma$, $x_\omega$ and $x_\rho$. In
the present work we use three parameterizations for the constants.
One of them is the standard parameterization GM1 given by
Glendenning--Moszkowski \cite{GM1}, as shown in Table I. This
parametrization employs ``low'' values for $x_\sigma$,  $x_\omega$
and  $x_\rho$. Larger values of these couplings make the EOS
stiffer and increase the value of the maximum mass of hadronic
stars to values above than 2 $M_{\odot}$, see Table I.


\section{The quark phase}
\label{qphase}

In order to study the just deconfined quark matter phase we use
an $SU(3)_f$ NJL effective model which
also includes color superconducting quark-quark interactions.
The corresponding lagrangian is given by
\begin{eqnarray}
{\cal L}
&=&
\bar \psi \left(i \rlap/\partial - \hat m \right) \psi
+ G \sum_{a=0}^8
\left[
\left( \bar \psi \ \tau_a \ \psi \right)^2
+ \left( \bar \psi \ i \gamma_5 \tau_a \ \psi \right)^2
\right]
\label{action}
\nonumber\\
& & \qquad + 2H
\sum_{A,A'=2,5,7}
\left[
\left(
\bar \psi \ i \gamma_5 \tau_A \lambda_{A'} \ \psi_C
\right)
\left(
\bar \psi_C \ i \gamma_5 \tau_A \lambda_{A'} \ \psi
\right)
\right]
\end{eqnarray}
where $\hat m=\mathrm{diag}(m_u,m_d,m_s)$ is the current mass matrix in
flavor space. The matrices, $\tau_i$ and $\lambda_i$ with $i=1,..,8$
are the
Gell-Mann matrices corresponding to the flavor and color groups respectively,
and $\tau_0 = \sqrt{2/3}\ 1_f$.
Finally, the charge conjugate spinors are defined as follows:
$\psi_C = C \ \bar \psi^T$ and $\bar \psi_C = \psi^T C$, where
$\bar \psi = \psi^\dagger \gamma^0$ is the Dirac conjugate spinor and
$C=i\gamma^2 \gamma^0$.

The lagrangian in Eq. (\ref{action}) leads to local chirally
invariant current-current interactions in the quark-antiquark and
quark-quark channels. The latter is expected to be responsible for
the presence of a color-superconducting phase in the region of low
temperatures and moderate chemical potentials. Here for simplicity
we do not include flavor mixing effects. The values of the quark
masses and the coupling constant $G$ can be obtained from the
meson properties in the vacuum. On the other hand, an estimate of $H/G$
can be obtained from Fierz transformation of
the one-gluon-exchange interactions in which case one gets $H/G =
0.75$, which is the value we will use hereafter.

To be able to determine the relevant thermodynamical quantities we
have to obtain the grand canonical thermodynamical potential
$\Omega(T,\mu_{fc})$ at finite temperature $T$ and chemical {\blue
potentials $\mu_{fc}$}. Here, $f=(u,d,s)$ and $c=(r,g,b)$ denote
flavor and color indices respectively. For this purpose we start
from Eq. (\ref{action}) and perform the usual bosonization of the
theory. This can be done by introducing scalar and pseudoscalar
meson fields $\sigma_a$ and $\pi_a$ respectively, together with
the bosonic diquark field $\Delta_A$. In this work we consider the
quantities obtained within the mean field approximation. Thus, we
only keep the non-vanishing vacuum expectation values of these
fields and drop the corresponding fluctuations. For the meson
fields this implies $\hat \sigma = \sigma_a\tau_a=
\textrm{diag}(\sigma_u,\sigma_d,\sigma_s)$ and $\pi_a=0$.
Concerning the diquark mean field, we will assume that in the
density region of interest only the 2SC phase might be relevant.
Thus, we adopt the ansatz $\Delta_5 = \Delta_7 = 0$, $\Delta_2 =
\Delta$. Integrating out the quark fields and working in the
framework of the Matsubara and Nambu-Gorkov formalism we obtain

\begin{eqnarray}
\Omega_\textrm{\tiny MFA}(T,\mu_{fc},\sigma_f,|\Delta|)
= -\frac{T}{2}\sum_{n=-\infty}^{\infty}\int\frac{d^3k}{(2\pi)^3}
\ln\det \left[\frac{S^{-1}(k)}{T}\right]
+
\frac{1}{4G}(\sigma_u^2+\sigma_d^2+\sigma_s^2)  + \frac{|\Delta|^2}{2H}
\end{eqnarray}
where
\begin{equation}
S^{-1}=\left(
\begin{array}{cc}
\slashed{k} - \hat M + \gamma_0 \hat\mu
&
-\Delta\gamma_5\tau_2\lambda_2
\\
\Delta^*\gamma_5\tau_2\lambda_2
&
\slashed{k} - \hat M - \gamma_0 \hat\mu
\end{array}
\right).
\end{equation}
Here, we have used $k = \left( (2n+1)\pi T , \vec k \right)$,
$\hat M =\textrm{diag}(M_u,M_d,M_s)$ with $M_f = m_f + \sigma_f$, and
$\hat\mu=\textrm{diag}(\{\mu_{fc}\})$ in flavor$\otimes$color space.
The determinant of this $72\times 72$ matrix can be
calculated analytically if $M_u=M_d$. Thus, in what follows
we will use the approximation $m_u=m_d \equiv m$
and $\sigma_u=\sigma_d\equiv\sigma$ which implies $M_u=M_d\equiv M$.
A detailed procedure for the calculation of the determinant can be
found in Ref.\cite{Huang:2002zd,Ruester:2005jc,Blaschke:2005uj}.
The resulting contribution to the thermodynamical potential is
\begin{eqnarray}
-\frac{T}{2}\sum_{n}\int\frac{d^3k}{(2\pi)^3} \ln\det [S^{-1}(k)/T]  =
2 \int^\Lambda \frac{d^3k}{(2\pi)^3} \sum_{\alpha=1}^9
\omega(x_i,y_i)
+ \mathrm{constant}
\end{eqnarray}
where $\Lambda$ is the cut-off of the theory and $\omega(x,y)$ is
defined by
\begin{eqnarray}
\omega(x,y)
= - \left[ x + T\ln[1+e^{-(x-y)/T}]
 + T\ln[1+e^{-(x+y)/T}] \right] \ ,
\end{eqnarray}
with
\begin{eqnarray}
x_{1,2} = E
\ \ , \ \
x_{3,4,5} = E_s
\ \ , \ \
x_{6,7} = \sqrt{\bigg[ E +  \frac{(\mu_{ur} \pm \mu_{dg})}{2} \bigg]^2 + \Delta^2}
\ \ , \ \
x_{8,9} = \sqrt{\bigg[ E +  \frac{(\mu_{ug} \pm \mu_{dr})}{2} \bigg]^2 + \Delta^2}
\nonumber
\end{eqnarray}
\begin{eqnarray}
y_1  = \mu_{ub}
\ , \ \
y_2 = \mu_{db}
\ , \ \
y_{3} = \mu_{sr}
\ , \ \
y_{4} = \mu_{sg}
\ , \ \
y_{5} = \mu_{sb}
\ , \ \
y_{6,7} = \frac{\mu_{ur}-\mu_{dg}}{2}
\ , \ \
y_{8,9} = \frac{\mu_{ug} - \mu_{dr}}{2}
\end{eqnarray}
where $E = \sqrt{ \vec k^2+ M^2}$, $E_s=\sqrt{ \vec k ^2 + M_s^2}$.

Finally, we include the thermodynamical potential for
non-interacting ultra-relativistic electrons
\begin{equation}
\Omega_e = -\frac{\mu_e^4}{12\pi^2}-\frac{\mu_e^2T^2}{6}-\frac{7\pi^2T^4}{180}, \nonumber
\end{equation}
where $\mu_e$ is the electron chemical potential. Therefore, the total potential for the quark matter
electron system is
\begin{eqnarray}
\Omega(T,\{\mu_{fc}\}, \mu_e, \sigma, \sigma_s)  & = &
\frac{1}{\pi^2}\int_0^\Lambda dk \; k^2 \sum_i \omega(x_i,y_i) +
\frac{1}{4G}(\sigma_u^2+\sigma_d^2+\sigma_s^2)  +
\frac{|\Delta|^2}{2H} - \Omega_\textrm{\tiny vac} + \Omega_e .
\end{eqnarray}
Here we have subtracted the constant $\Omega_\textrm{\tiny vac}$
in order to have a vanishing pressure at vanishing temperature and
chemical potentials. From the grand thermodynamic potential
$\Omega$ we can readily obtain the pressure $P = - \Omega$, the
number density of quarks of each flavor and color
\begin{equation}
n_{fc} =  - \frac{ \partial \Omega }{\partial \mu_{fc} }  =
-\frac{1}{\pi^2}\int_0^\Lambda dk \; k^2
\frac{\partial}{\partial \mu_{fc}} \bigg( \sum_i \omega(x_i,y_i)  \bigg) ,
\end{equation}
and the number density of electrons
\begin{eqnarray}
n_{e} = - \frac{ \partial \Omega }{\partial \mu_e}.
\end{eqnarray}
The corresponding number densities of each flavor, $n_f$, and of
each color, $n_c$, in the quark phase are given by
\begin{eqnarray}
n_f = \sum_{c} n_{fc} \qquad , \qquad n_c = \sum_{f} n_{fc}.
\end{eqnarray}
Finally, the baryon number density $n_B$ reads
\begin{equation}
n_B = \frac{1}{3} \sum_{fc} n_{fc} = \frac{1}{3} (n_u + n_d + n_s)
\end{equation}
and the Gibbs free energy per baryon is
\begin{equation}
g_\textrm{\scriptsize quark}= \frac{1}{n_B}\left(\sum_{fc} \mu_{fc} \ n_{fc}+ \mu_e \ n_e \right).
\label{g_quark}
\end{equation}

For the NJL model we use two sets of parameters.
They have been derived from those used in refs. \cite{Rehberg:1995kh}
and \cite{Hatsuda:1994pi} by neglecting the 't Hooft interaction.
To do that we have followed the procedure proposed in
\cite{Buballa2005}.
Namely, keeping $\Lambda$ and $m$ fixed we have varied $G$ and $m_s$ in
order to obtain $M =367.6$ MeV and $M_s=549.5$ MeV at zero temperature
and density.
The resulting parameter sets are given in Table \ref{sets}.

\begin{table*}[t!]
\centering
\begin{tabular}{c|ccccc}
\hline\hline
& $m_{u,d} $ [Mev] & $m_s$ [Mev] & $\Lambda$ [Mev] &  $G\Lambda^2$&
$H/G$\\
\hline
set 1 & 5.5 & 112.0  & 602.3 & 4.638 & 3/4 \\
set 2 & 5.5 & 110.05 & 631.4 & 4.370 & 3/4 \\
\hline
\end{tabular}
\caption{The two sets of NJL parameters.}
\label{sets}
\end{table*}


In order to derive a quark matter EOS from the above formulae it
is necessary to impose a suitable number of conditions on the
variables  $\{\mu_{fc}\}, \mu_e, \sigma, \sigma_s$ and $\Delta$.
Three of these conditions are consequences from the fact that the
thermodynamically consistent solutions correspond to the
stationary points of $\Omega$ with respect to $\sigma$,
$\sigma_s$, and $\Delta$. Thus, we have
\begin{eqnarray}
\partial\Omega/\partial\sigma =0
\qquad , \qquad
\partial\Omega/\partial\sigma_s =0
\qquad , \qquad \partial\Omega/\partial|\Delta|=0.
\label{gapeq}
\end{eqnarray}
To obtain the remaining conditions one must specify the physical
situation in which one is interested in. In many astrophysical
applications considered in the literature quark matter in
$\beta$-equilibrium was analyzed. In such a case chemical
equilibrium is maintained by weak interactions among quarks, e.g.
$d \leftrightarrow u + e^- + \bar{\nu}_e$, $s \leftrightarrow u +
e^- + \bar{\nu}_e$, $u + d \leftrightarrow u + s$. Moreover, for
neutron stars older than a few minutes, neutrinos can leave the
system. Thus, lepton number is not conserved and we have four
independent conserved charges, namely the electric charge $n_Q =
\frac{2}{3} n_u - \frac{1}{3} n_d - \frac{1}{3} n_s - n_e$ and the
three color charges $n_u$, $n_d$ and $n_s$. Instead of $n_u$,
$n_d$ and $n_s$, the linear combinations $n = n_r + n_g + n_b$,
$n_3 = n_r - n_g$ and $n_8 = \frac{1}{\sqrt{3}}  (n_r + n_g - 2
n_b)$ are often used. Here $n$ is the total quark number density
(i.e. $n = 3 n_B$) and  $n_3$ and $n_8$ describe color
asymmetries. The four conserved charges $\{n_j\} = \{n, n_3, n_8,
n_Q\}$ are related to four independent chemical potentials
$\{\mu_j\} =\{\mu, \mu_3, \mu_8, \mu_Q\}$ such that $n_j = - {
\partial \Omega }/{\partial \mu_j}$. The individual quark chemical
potentials $\mu_{fc}$ are given by
\begin{eqnarray}
\mu_{fc} &=& \mu + \mu_Q \left[ \frac{1}{2} (\tau_3)_{ff} +
\frac{1}{2\sqrt{3}} (\tau_8)_{ff} \right]
+ \mu_3 (\lambda_3)_{cc}  + \mu_8 (\lambda_8)_{cc}
\end{eqnarray}
where, as before, $\tau_i$  and $\lambda_i$ are the Gell-Mann
matrices in flavor and color space respectively. The electron
chemical potential is $\mu_e = - \mu_Q$, thus we have
\begin{eqnarray}
\mu_{dc} = \mu_{sc} = \mu_{uc} + \mu_e
\label{beta}
\end{eqnarray}
for all colors $c$, which are the $\beta$-equilibrium conditions. For electrically and
color neutral matter we have also the conditions:
\begin{eqnarray}
n_Q \equiv
- \frac{ \partial \Omega }{\partial \mu_Q} = 0
\qquad , \qquad
n_3 \equiv
- \frac{ \partial \Omega }{\partial \mu_3} = 0
\qquad , \qquad
n_8 \equiv
- \frac{ \partial \Omega }{\partial \mu_8} = 0 \; .
\label{colden}
\end{eqnarray}
Employing the above conditions, the system can be characterized by
two independent variables, e.g. $(T,n_B)$ or $(T,P)$.


The conditions given in Eqs.(\ref{beta},\ref{colden}) are extensively
employed to describe quark matter
in $\beta$-equilibrium.
However, in this paper we deal with \emph{just} deconfined quark matter which is
temporarily out of $\beta$-equilibrium.
As already emphasized in 
\cite{Madsen1994,IidaSato1998,Lugones1998,Lugones1999,Bombaci2004,Lugones2005,Bombaci2007}
the appropriate condition in this case is flavor
conservation between hadronic and deconfined quark matter. This
can be written as
\begin{equation}
Y^H_f = Y^Q_f   \;\;\;\;\;\; f=u,d,s,L \label{flavor}
\end{equation}
being $Y^H_f \equiv n^H_f / n^H_B$ and  $Y^Q_i \equiv
n^Q_f / n^Q_B$ the abundances of each particle in the hadron and
quark phase respectively. In other words, the just deconfined quark
phase must have the same ``flavor'' composition than the
$\beta$-stable hadronic phase from which it has been originated. Notice that,
since the hadronic phase is assumed to be electrically neutral, flavor conservation
ensures automatically the charge neutrality of the just deconfined quark phase.

The conditions given in Eq. (\ref{flavor}) can be
re-written as follows
\begin{equation}
n_d = \xi ~ n_u   \label{h1},
\end{equation}
\begin{equation}
n_s = \eta ~ n_u   \label{h2},
\end{equation}
\begin{equation}
3 n_{e} = 2 n_{u} - n_{d} - n_{s} ,\label{h3}
\end{equation}
\noindent where $n_i$ is the particle number density of the
$i$-species in the quark phase. The quantities $\xi \equiv Y^H_d /
Y^H_u$ and $\eta \equiv Y^H_s / Y^H_u$ are functions of the
pressure and temperature, and they characterize the composition of
the hadronic phase. These expressions are valid for \textit{any}
hadronic EOS. For hadronic matter containing $n$, $p$, $\Lambda$,
$\Sigma^{+}$, $\Sigma^{0}$, $\Sigma^{-}$, $\Xi^{-}$ and $\Xi^{0}$,
we have
\begin{eqnarray}
\xi &=& \frac{n_p  +  2  n_n  + n_{\Lambda} + n_{\Sigma^{0}} +  2
n_{\Sigma^{-}}  + n_{\Xi^{-}}}{2  n_p  +  n_n  +  n_{\Lambda} + 2
n_{\Sigma^{+}} + n_{\Sigma^{0}}  +  n_{\Xi^{0}}}, \label{xi} \\
\eta &=& \frac{n_{\Lambda}  + n_{\Sigma^{+}} + n_{\Sigma^{0}}  +
n_{\Sigma^{-}} + 2 n_{\Xi^{0}} + 2 n_{\Xi^{-}}}{2  n_p  +  n_n  +
n_{\Lambda} + 2 n_{\Sigma^{+}} + n_{\Sigma^{0}}  +  n_{\Xi^{0}}}.  \label{eta}
\end{eqnarray}
Additionally, the deconfined phase must be locally colorless; thus
it must be composed by an equal number of red, green and blue
quarks
\begin{equation}
n_r = n_g = n_b
 \label{colorless}.
\end{equation}
Also, $ur$, $ug$, $dr$, and $dg$ pairing will happen provided that $|\Delta|$ is nonzero, leading to
\begin{equation}
 n_{ur}=n_{dg},\quad n_{ug}=n_{dr}.
\label{pairing}
\end{equation}

In order to have all Fermi levels at the same value, we consider \cite{Lugones2005}
\begin{eqnarray}
 n_{ug} &=& n_{ur}\\
 n_{sb} &=& n_{sr}.
\label{equalfermilevels}
\end{eqnarray}
These two equations, together with Eqs. (\ref{colorless})
and (\ref{pairing}) imply that $n_{ur}=n_{ug}=n_{dr}=n_{dg}$ and
$n_{sr}=n_{sg}=n_{sb}$.

Finally, including the conditions given in Eq.(\ref{gapeq}) we have 12
equations involving the  13 unknowns ($\sigma$, $\sigma_s$,
$|\Delta|$, $\mu_e$ and $\{\mu_{fc}\}$). For given value of one of
the chemical potentials (e.g. $\mu_{ur}$), the set of equations
can be solved once the values of the parameters $\xi$,  $\eta$ and
the temperature $T$ are given. Instead of $\mu_{ur}$, we can
provide a value of the Gibbs free energy per baryon
$g_\textrm{\scriptsize quark}$ or the pressure $P$ and solve
simultaneously Eqs. (\ref{h1})-(\ref{equalfermilevels}) together
with Eq. (\ref{gapeq}) in order to obtain $\sigma$, $\sigma_s$,
$|\Delta|$, $\mu_e$ and $\{\mu_{fc}\}$.

The above conditions represent a state that fulfills all
the physical requirements of the just deconfined phase, i.e. it is color
and electrically neutral, and it has a ``flavor content" determined by
the parameters $\xi$ and $\eta$ (both related to the hadronic phase
through Eqs. (\ref{xi}) and (\ref{eta})).

The set of twelve equations can be summarized as follows \cite{Lugones2005}
\begin{eqnarray}
n_{ub} = 2\ \frac{2-\xi}{1+\xi}\ n_{ur},
\ \   \ \
n_{db} = 2\ \frac{2\xi-1}{1+\xi}\ n_{ur},
\ \  \ \
n_{sr} = n_{sg} =n_{sb} =  \frac{2 \eta }{1+\xi}\ n_{ur},
\nonumber
\end{eqnarray}
\begin{eqnarray}
n_{ug} = n_{dr} = n_{dg} = n_{ur},
\ \  \ \
n_e  =   \frac{2(2 - \xi - \eta)}{1 + \xi} \ n_{ur},
\end{eqnarray}
together with equation (\ref{gapeq}).

\section{Deconfinement transition in neutron star matter}

\label{disc}

\begin{figure}
 \includegraphics[scale=.4]{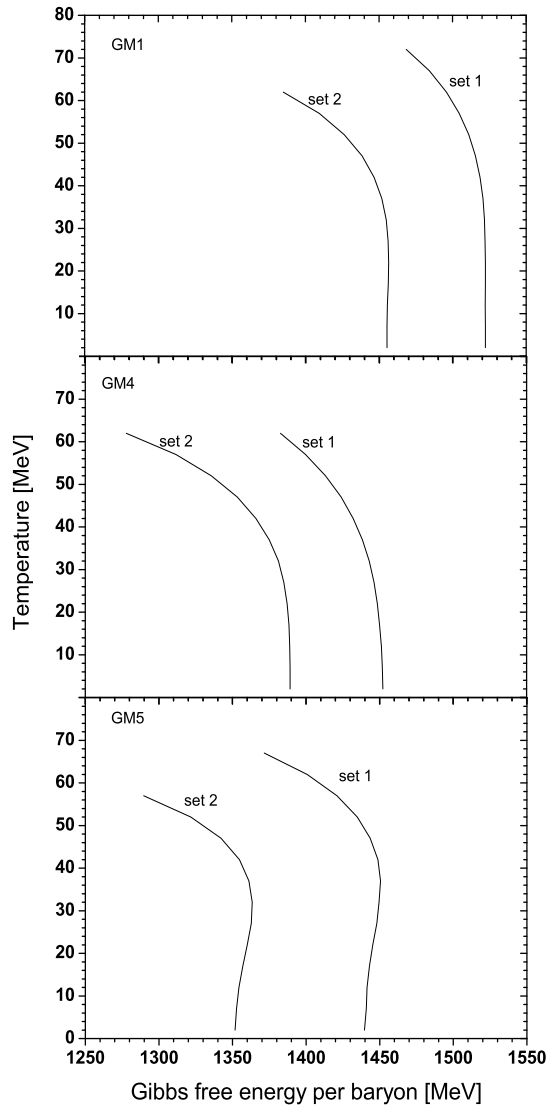}
\caption{For each temperature we show the Gibbs free energy density per
barion $g$ of the deconfinement phase transition. The hadronic phase is
described by the GM1, GM4 and GM5 parametrization of the EOS given in
Sec. II (see Table I). For the quark phase we adopt the two parameterizations
of the NJL model given in Table II. If hadronic matter has a temperature $T_h$
and a Gibbs free energy density per barion $g_h$ lying to the left of a
given curve, then the deconfinement transition is not possible for that
parameterizations of the EOSs. If the point $(g_h,T_h)$ lies to the
right of the curve the preferred phase is deconfined quark matter.}
\label{Tg}
\end{figure}

For simplicity, the analysis that follows will be made in bulk,
i.e. without taking into account the energy cost due to
finite size effects in creating a drop of deconfined quark
matter in the hadronic environment. Finite size effects
on the nucleation of color superconducting quark bubbles in a
cold deleptonized hadronic medium have been recently analyzed employing
the MIT bag model in the description of the quark phase  \cite{Bombaci2007}.
As a consequence of the surface effects, it is necessary to have an
overpressure with respect to the bulk transition point. However, since this
effect is not very large we leave its consideration for a future more detailed work.

In order to determine the transition conditions, we apply the
Gibbs criteria, i.e. we assume that deconfinement will occur when
the pressure and Gibbs energy per baryon are the same for both
hadronic matter and quark matter at a given common temperature.
Thus, we have
\begin{eqnarray}
g_h = g_q
\qquad , \qquad
P_h = P_q
\qquad , \qquad
T_h = T_q \;  ,
\end{eqnarray}
where the index $h$ refers to hadron matter and the index $q$ to quark matter.
The results are displayed in Fig. \ref{Tg} where we show the Gibbs free
energy per baryon $g$ at the transition point as a function of the temperature. The hadronic
phase is described by the three parameterizations of the EOS given in Sec. II (see Table I).
For the quark phase we adopt the NJL model described above (using flavor conservation
conditions) with the two parameterizations given in Table II.
If hadronic matter has a temperature $T_h$ and a Gibbs free energy per
baryon $g_h$ lying to the left of a given curve, then the deconfinement
transition
is not possible for that parameterizations of the EOSs. In the
right side region of
a given curve the preferred phase is deconfined quark matter.


Let us now examine the consequences of the above results on the structure of compact stars.
Stars containing quark phases fall into two main classes: hybrid
stars (where quark matter is restricted to the core) and strange
stars (made up completely by quark matter). It is expected that both kinds
of stars cannot exist simultaneously in Nature, but it is not know which one
would be realized (if any).  This depends on whether the energy per baryon of
$\beta$-equilibrated quark matter at zero pressure and zero
temperature is less than the neutron mass (the so called ``absolute
stability'' condition \cite{fj84}). Analysis made within the MIT bag model shows that
there is a room in the parameter space for the existence of strange
stars. Moreover,
pairing enlarges substantially the region of the parameter space
where $\beta$-stable quark matter has an energy per baryon smaller
than the neutron mass \cite{Lugones2002,Lugones2003}. Although the
gap effect does not dominate the energetics, being of the order
($\Delta / \mu)^2 \sim$ a few percent, the effect is substantially
large near the zero-pressure point (which determines the stability
and also the properties of the outer layers and surface of the
star). As a consequence, a ``CFL strange matter'' is allowed for the
same parameters that would otherwise produce unbound strange matter
without pairing \cite{Lugones2002}. However, within the NJL model, the strange matter
hypothesis is not favored, at least for the most accepted parameterizations
of the EOS \cite{Buballa2005}. Thus, stars containing quark phases are believed to be
hybrid stars within the NJL model.

\begin{figure}
 \includegraphics[scale=.3]{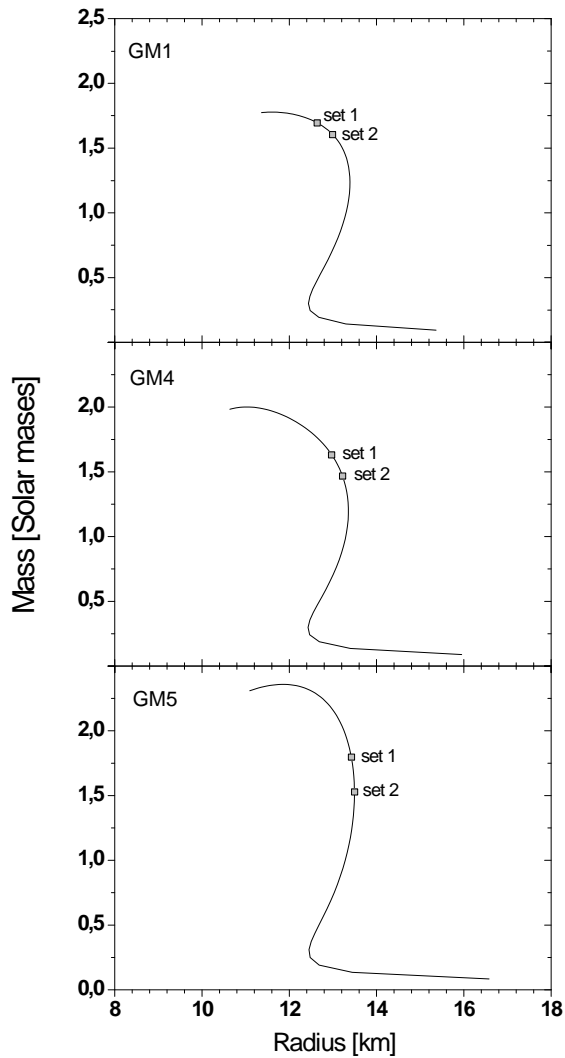}
\caption{Mass-radius relationship for hadronic stars at zero temperature
with the GM1, GM4 and GM5 EOS.
For a given parametrization of the NJL model, stars with a mass above
the corresponding
point given in the figure are hybrid stars containing quark cores. The
mass-radius relationship
for these hybrid stars is \textit{not} shown in this figure.}
\label{MR}
\end{figure}

At a given temperature there is a univocal relation between the
mass of a compact star and its central pressure (or equivalently
the Gibbs free energy per barion at the center of the star). Thus,
we can employ the results given in Fig. \ref{Tg} to calculate the
\textit{critical} compact star mass $M_{cr}$ above which they
should contain a quark core. With this purpose, we integrate the
Tolman-Oppenheimer-Volkoff equations of relativistic stellar
structure employing the hadronic EOSs given in Table I, and
identify the mass of the \textit{pure hadronic star} for which the
Gibbs free energy per barion $g_h$ at the center  is equal to the
\textit{critical}  $g$ given in Fig. \ref{Tg}. This is called
\textit{critical} mass $M_{cr}$ because  pure hadronic stars with
$M_{h}
> M_{cr}$ are very unlikely to be observed, while pure hadronic
stars with $M_{h} < M_{cr}$ are safe with respect to a sudden
transition to quark matter.
The results are shown in Fig. \ref{MR} for neutron stars at zero
temperature where we show the mass-radius relationship for hadronic stars and indicate the
critical mass for the two selected parameterizations of the NJL model.

In the first seconds  after their formation in a core collapse supernova explosion,
neutron stars may have temperatures up to $\sim$ 50 MeV and a large amount of trapped
neutrinos. Thus, the results presented in Fig. \ref{Tg} are not appropriate
for the analysis of the critical mass in proto-neutron star conditions since they do
not include the effect of trapped neutrinos.
However, trapped neutrinos increase considerably the critical density
$\rho_{cr}$ for the transition to quark matter \cite{Lugones1998}. Thus, it
is possible that the transition is strongly inhibited in the initial moments
of the evolution of neutron stars \cite{Lugones1999}.
About one minute after its birth, there are almost no more trapped neutrinos
in neutron star matter, while the temperature is still high
(up to $\sim 10$ MeV \cite{Pons1999}). Thus, the results presented in Fig. \ref{Tg} are
valid for neutron stars older than $\sim$ 1 minute.
On the other hand, as apparent from Fig. \ref{Tg}, there is little variation in
the \textit{critical} Gibbs free energy per barion for temperatures below $\sim 10$ MeV.
Thus, the critical masses presented in Fig. \ref{MR} are also valid for neutron stars older
than $\sim$ 1 minute (in spite of being calculated considering hadronic matter at zero temperature).

\section{Discussion and Conclusions}
\label{summary}

In this paper we have analyzed the deconfinement transition from
hadronic matter to quark matter and investigated the role of color
superconductivity within the Nambu-Jona-Lasinio model. The  study
presented here is relevant for neutron stars older than $\sim 1$
minute, when there are almost no more trapped neutrinos in neutron
star matter.

For the hadronic phase we have used three different
parameterizations of a non-linear Walecka model which includes the
whole baryon octet and electrons. One of them is the standard
parametrization GM1 given by Glendenning--Moszkowski \cite{GM1}.
This parametrization employs ``low'' values for the relative
meson-hyperon couplings $x_{\sigma}$,  $x_{\omega}$ and
$x_{\rho}$. Larger values of these couplings make the EOS stiffer
and, as shown in Table I, increase the value of the maximum mass of
hadronic stars to values above 2 $M_{\odot}$. These large
values may be relevant in connection with recent measurements of
highly-massive neutron stars which give one of the most stringent
test on the overall stiffness of dense matter EOS. However, such
measurements still have to be taken with caution (see discussion
given in \cite{Bielich2008}). For example, the mass of the pulsar
J0751+1807 was corrected from $M=(2.1\pm 0.1)M_\odot$
\cite{Nice:2005fi} to $(1.26\pm 0.14)M_\odot$ as new data became
available \cite{Nice:2008}.
There is also a series of measurements of
extremely massive pulsars in globular clusters, where just the
periastron advance has been determined but not the inclination angle
of the orbit \cite{Ransom:2005ae,frei07b,frei07a}. For the
pulsar PSR J1748-2021B a mass of $(2.74\pm 0.21)M_\odot$ is reported
by using a statistical analysis for the inclination angle
\cite{frei07b}.
Also, recent measurements of post-Keplerian orbital parameters
in relativistic binary systems containing millisecond pulsars
give evidence for the existence of highly-massive
compact stars.  For example, the compact star associated to the millisecond pulsar
PSR~B1516+02B in the Globular Cluster NGC~5904 (M5) has a mass
$M = 1.94^{+ 0.17}_{- 0.19} M_{\odot}$  ($1~\sigma$) \cite{frei07a}.
Other constraints for the mass and radius have been obtained studying
redshifted spectral lines extracted in the aftermath of an X-ray burst
of  the Low Mass X-ray Binary EXO 0748-676 in 2000 \cite{Cottam:2002cu}.
A model analysis of the X-ray burst led to
rather tight constraints for the mass and radius of the compact star
of $M\geq (2.10\pm 0.28)M_\odot$ and $R\geq (13.8\pm 1.8)$~km
\cite{Ozel:2006km} which are based on the redshift measurement of
\cite{Cottam:2002cu}.  However, a detailed multiwavelength analysis
concluded that the mass of the compact star is more compatible with
$1.35M_\odot$ than with $2.1M_\odot$ \cite{Pearson:2006zy}.
Moreover, follow-up observations of another burst in 2003
\cite{Cottam:2007cd} could not confirm
the spectral features seen in the burst spectra presented in \cite{Cottam:2002cu}.
While most of these measurements would need further confirmation,
it is worth also exploring hadronic models that can produce stellar configurations
with masses above 2 $M_{\odot}$ (see the parameterizations GM4 and GM5 in Table I).

Employing the results of Fig. 1 and integrating the Tolman-Oppenheimer-Volkoff
equations of relativistic stellar structure for the hadronic EOSs given in Table I
we have calculated the \textit{critical} compact star mass $M_{cr}$ above which they should contain a quark core.
Pure hadronic stars with $M_{h} > M_{cr}$ are very unlikely to exist, while pure hadronic
stars with $M_{h} < M_{cr}$ are safe with respect to a sudden transition to quark matter.
Notice that the critical mass is defined here in a different way as
in Refs. \cite{Bombaci2004,Bombaci2007,Bombaci2008}. In the here-presented
\textit{bulk} analysis, the transition begins when
the Gibbs conditions  $\Delta P \equiv P_h - P_q = 0$ and $\Delta g \equiv g_h - g_q=  0$
are verified at a given temperature. However, for the nucleation of finite size bubbles,
it is necessary to have an overpressure $\Delta P > 0$ with respect to the bulk transition
point due to surface and curvature effects. For a given overpressure there is a
probability (and a corresponding nucleation time) to nucleate a quark bubble due to quantum
or thermal fluctuations.
In \cite{Bombaci2004,Bombaci2007} the critical mass was defined as the value of the gravitational mass of a
hadronic star for which the nucleation time due to quantum fluctuations is equal to one year.
Calculations of the critical mass done within the frame of the MIT bag model show that surface effects
are strong for values of the Bag constant $B$ smaller than $\sim ~ 100 ~ \mathrm{MeV ~ fm^{-3}}$ \cite{Bombaci2007}.
That is, within the MIT bag model, the effect of the surface tension $\sigma$ is strong for strange stars but
it is  small if the parameters of the EOS correspond to hybrid stars.
For the parameterizations of the NJL model employed in this work only hybrid stars are allowed.
Thus, finite size effects are not expected to introduce qualitative modifications in our results.

Employing set 1 for the Nambu-Jona-Lasinio model (see Table II) we
find that there are not large variations in the critical mass for
the three different parameterizations of the hadronic matter
equation of state. As seen in Fig. 2 the critical mass is in the
range 1.65-1.80 $M_{\odot}$ for set 1, which is not a large
difference if we consider the larger variation in the maximum mass
of hadronic stars within the three parameterizations ($M_{max}$
between 1.78 and 2.35 $M_{\odot}$, see Table I). A similar result
is found employing set 2 for the Nambu-Jona-Lasinio model, i.e.
the  critical mass is in the range 1.45-1.60 $M_{\odot}$ (see Fig.
2). This is in coincidence with previous results obtained for
hybrid stars within the MIT Bag model \cite{Bombaci2007}. As
mentioned above, in the case of the MIT Bag model stars containing
quark phases are strange stars for low values of the Bag constant
$B$, and hybrid stars for sufficiently large values of $B$. For
the values of $B$ corresponding to strange stars the critical mass
may vary essentially from zero to near the maximum mass of
hadronic stars, depending on the value of other parameters such as
the superconducting gap $\Delta$ and the surface tension $\sigma$.
However, for the values of $B$ corresponding to hybrid stars the
critical mass is always close to (but smaller than) the maximum
mass of hadronic stars, rather independently of the value of other
parameters \cite{Bombaci2007}. The parameterizations employed here
of the Nambu-Jona-Lasinio model allow only for the existence of
hybrid stars and show the same characteristics of the critical
mass, allowing the existence of a mixed population of compact
stars (pure hadronic up to the critical mass and hybrid above the
critical mass).

\section{Acknowledgements}

This work was supported in part by CONICET (Argentina) grant \# PIP 6084 and
by ANPCyT  (Argentina) grants \# PICT04 03-25374 and \# PICT07 03-00818.

\end{document}